\documentclass[12pt]{article}

\begin{document}
\begin{center}
{\bf Quantization of Fermionic Fields with Two Mass States in the First Order Formalism }\\
\vspace{5mm}
 S. I. Kruglov \\
\vspace{5mm}
\textit{University of Toronto at Scarborough,\\ Physical and Environmental Sciences Department, \\
1265 Military Trail, Toronto, Ontario, Canada M1C 1A4}
\end{center}

\begin{abstract}
The relativistic 20-component wave equation, describing particles
with spin 1/2 and two mass states, is analyzed. The projection
operators extracting states with definite energy and spin
projections, and density matrix are obtained. The canonical
quantization of the field with two mass states in the formalism of
the first order is performed and the chronological pairing of the
20-component operators was found.

\end{abstract}

\section{Introduction}

It is known that the problem of quark and lepton generations, and
their mass spectrum has not been solved yet. The standard model
(SM) of electroweak interactions does not explain the number of
quark and lepton generations, mass spectrum and contains many free
parameters. Barut suggested \cite{Barut1}, (see also
\cite{Barut2}, \cite{Barut3}, \cite{Wilson}) to describe $e$-,
$\mu$-leptons by the generalized Dirac equation of the second
order. He interpreted this equation, based on the non-perturbative
approach to quantum electrodynamics (QED), as an effective
equation for partly ``dressed" fermions. Some investigations of
the generalized Dirac equation of the second order were made in
\cite{Dvoeglazov} We have formulated this equation in the form of
the first order generalized Dirac equation (FOGDE) \cite{Kruglov}.
This form is convenient for analyzing symmetry properties of the
theory and different calculations. Here we continue to explore
this matrix form of the equation.

The paper is organized as follows. In Sec. 2, we formulate FOGDE
and corresponding Lagrangian. The projection operators extracting
solutions with definite energy and spin projections for free
particles are given in Sec. 3. In Sec. 4, we perform the canonical
quantization of fields. The commutation relations for 20-component
functions and the chronological pairing of operators were
obtained. We make a conclusion in Sec. 6. Appendix contains some
useful formulas. We use notations as in \cite{Ahieser} and the
system of units $\hbar =c=1$ is chosen.

\section{First Order Field Equation}

The Barut second order generalized Dirac equations \cite{Barut1},
describing spin-1/2 particles, which possess two mass states, can
be represented in the form \cite{Kruglov}:
\begin{equation}
\left(\gamma_\nu\partial_\nu- \frac{a}{m}\partial_\mu^2 +
m\right)\psi(x)=0 . \label{1}
\end{equation}
where $\partial_\nu =\partial/\partial x_\nu =(\partial/\partial
x_m,\partial/\partial (it))$, $\psi (x)$ is a Dirac spinor and the
Dirac matrices $\gamma_\mu $ obey the commutation relations
$\gamma_\mu \gamma_\nu +\gamma_\nu \gamma_\mu =2\delta_{\mu\nu}$.
The $m$ is a parameter with the dimension of the mass, and $a$ is
a massless parameter in Eq. (1).

Masses of spin-1/2 particles are given by
\begin{equation}
m_1=\pm m\left(\frac{1 - \sqrt{4a+1}}{2a}\right) ,~~m_2=\pm
m\left(\frac{1 +\sqrt{4a+1}}{2a}\right) .
 \label{2}
\end{equation}
For a given parameter $a$, one can chose the positive values (for
$a\geq -1/4$) of masses $m_1$, $m_2$ in Eq. (2).

According to Barut's idea \cite{Barut1}, the states of fermions
with two masses are interpreted as an electron and a muon. One can
apply Eq. (1) for describing quarks or massive neutrinos.

It was shown in \cite{Kruglov} that if we introduce the
20-component function
\begin{equation}
\Psi (x)=\left\{ \psi _A(x)\right\} =\left(
\begin{array}{c}
\psi (x)\\
\psi _\mu (x)
\end{array}
\right) , \label{3}
\end{equation}
where
\[
  \psi_\mu (x)=-\frac{1}{m}\partial_\mu \psi (x)
\]
and index $A$ runs values $A=0,\mu$, Eq. (1) becomes
\begin{equation}
\left( \alpha _\nu \partial _\nu +m\right) \Psi (x)=0 . \label{4}
\end{equation}
The 20-dimensional matrices of the first order wave equation (4)
are
\begin{equation}
\alpha _\nu =\left(\varepsilon ^{\nu,0 }+ a\varepsilon ^{0,\nu
}\right)\otimes I_4 + \varepsilon^{0,0}\otimes\gamma_\nu ,
\label{5}
\end{equation}
where $I_4$ is the unit $4\times 4$ matrix, and $\otimes$ means
the direct product of matrices. The elements of the entire matrix
algebra $\varepsilon ^{A,B}$ \cite{Kruglov1} obey equations as
follows:
\begin{equation}
\left( \varepsilon ^{M,N}\right) _{AB}=\delta _{MA}\delta _{N,B},
\hspace{0.5in}\varepsilon ^{M,A}\varepsilon ^{B,N}=\delta
_{AB}\varepsilon ^{M,N}, \label{6}
\end{equation}
where $A,B,M,N=0,1,2,3,4$. The first equation in (6) shows the
matrix elements of the matrix $\varepsilon ^{M,N}$, and the second
equation presents the product of two matrices. The relativistic
wave equation of the first order (4) is convenient for different
applications because the matrices of the equation, $\alpha _\nu$,
expressed through the elements of the entire matrix algebra with
simple properties (6).

The Lagrangian is given by
\begin{equation}
{\cal L}=-\frac{1}{2}\left[\overline{\Psi }(x)\alpha _\mu
\partial _\mu \Psi (x)-\left(\partial _\mu\overline{\Psi }(x)\right)\alpha _\mu
\right] -m \overline{\Psi }(x)\Psi (x) ,\label{7}
\end{equation}
where $\overline{\Psi }(x)=\Psi ^{+}(x)\eta$, $\eta$ is the
Hermitianizing matrix, and $\Psi ^{+}(x)$ is the
Hermitian-conjugate wave function. The Euler-Lagrange equation (4)
follows from the Lagrangian (7). The Hermitianizing matrix $\eta$
obeys the relations \cite{Ahieser}
\begin{equation}
\eta \alpha _m=-\alpha _m^{+}\eta^+ ,~~~\eta \alpha _4=\alpha
_4^{+}\eta^+ ,~~~ \eta^+=\eta~~~(m=1,2,3) .  \label{8}
\end{equation}
and is given by \cite{Kruglov}
\begin{equation}
\eta=\left(a\varepsilon^{m,m}-a\varepsilon^{4,4}
-\varepsilon^{0,0} \right)\otimes\gamma_4 . \label{9}
\end{equation}
Eq. (8) guarantees that the Lagrangian (7) is the real function.
By varying the action $S=\int {\cal L}d^3x$ with respect to
$\Psi(x)$, one may obtain the second Euler-Lagrange equation:
\[
\left(\partial _\nu \overline{\Psi} (x)\right) \alpha _\nu -
 m \overline{\Psi} (x)=0 .
\]
This equation may also be obtained by Hermitian conjugating Eq.
(4), and multiplying it on $\eta$, and taking into account Eqs.
(8).

\section{ Mass and Spin Projection Operators}

Now we consider states of particles with definite energy, $p_0$,
and momentum, $\textbf{p}$. It should be noted that for definite
mass state, one has
\begin{equation}
p_0=\sqrt{\textbf{p}^2 +m_1^2},~~~ or ~~~ p_0=\sqrt{\textbf{p}^2
+m_2^2}. \label{10}
\end{equation}
This means that there is an additional quantum number $\tau$, so
that $\tau=1$, $2$ for states with masses $m_1$ and $m_2$
\cite{Wilson}. We omit this index $\tau$ in four momentum
$p=(\textbf{p},ip_0)$ keeping this in mind. Solutions of Eq. (4)
with definite energy and momentum in the form of plane waves are
given by
\begin{equation}
\Psi^{(\pm)}_{\pm p}(x)=\frac{1}{\sqrt{2p_0 V}}U(\pm p)\exp(\pm
ipx) , \label{11}
\end{equation}
where $V$ is the normalization volume. There are two solutions
(11) corresponding to Eqs. (10). Replacing Eq. (11) into Eg. (4),
one obtains
\begin{equation}
\left(\pm i \check{p}+ m \right)U(\pm p)=0 , \label{12}
\end{equation}
where $\check{p}=\alpha_\mu p_\mu$. The solution $\Psi^{(+)}_{+
p}(x)$ with positive frequency $\omega=p_0$ describes particles
and the solution $\Psi^{(-)}_{- p}(x)$ with negative frequency
$\omega=-p_0$ corresponds to antiparticles. We use here the
normalization condition
\begin{equation}
\int_V \overline{\Psi}^{(\pm)}_{\pm p}(x)\alpha_4
\Psi^{(\pm)}_{\pm p}(x)d^3 x=1 , \label{13}
\end{equation}
where $\overline{\Psi}^{(\pm)}_{\pm p}(x)=\left(\Psi^{(\pm)}_{\pm
p}(x)\right)^+ \eta$. It is implied the integration over the
volume $V$ in Eq. (13). The difference from QED is that all
functions in Eqs. (11)-(13) are 20-component functions and
$\eta\neq \alpha_4$. In addition, Eqs. (11)-(13) are valid for two
mass values ($\tau =1,2)$, Eq. (10). Normalization conditions for
20-component functions $U(\pm p)$ follow from Eqs. (11)-(13):
\begin{equation}
\overline{U}(\pm p)\alpha_\mu U(\pm p)=-2ip_\mu , \label{14}
\end{equation}
\begin{equation}
\overline{U}(\pm p)U(\pm p)=\mp 2\frac{p^2}{m} . \label{15}
\end{equation}
For $\tau=1$, $p^2=-m_1^2$, and for $\tau=2$, $p^2=-m_2^2$. Eqs.
(14), (15) are analogs of normalization conditions for bispinors
in QED. But, using the definition of the 20-component function
$\Psi (x)$, Eq. (3), we find, from Eq. (15), the normalization
condition for bispinor $u(\pm p)$ (see also \cite {Wilson} for
comparison):
\begin{equation}
\overline{u}(\pm p)u(\pm p)=\pm  \frac{2mp^2}{m^2 -ap^2} ,
\label{16}
\end{equation}
where $\overline{u}(\pm p)=u^+(\pm p)\gamma_4$. The normalization
condition (16) for bispinors is different from QED. For the case
of one mass state, $a=0$, $p^2=-m^2$ (see Eq. (1)), we come to
QED, and Eq. (16) becomes the standard normalization condition
\cite{Ahieser}\footnote{To have the same normalization condition
as in QED for $a=0$, $p^2=-m^2$, one should make the replacement
$\eta\rightarrow -\eta$, which does not change Eqs. (8)}.

Now we construct the projection matrix extracting solutions of Eq.
(12) corresponding to definite energy and momentum. With the help
of Eqs (5), (6) (see Appendix), we obtain the minimal equation for
the matrix $\check{p}=p_\mu \alpha_\mu$:
\begin{equation}
\check{p}^5 -\left(1+2a\right)p^2 \check{p}^3 + a^2 p^4
\check{p}=0 . \label{17}
\end{equation}
Using Eq. (17) it is not difficult to verify that the matrices
\begin{equation}
\Pi_{\pm}=\frac{\pm i\check{p}\left(m\mp i\check{p}\right)
\left[\check{p}^2-\left(1+2a\right)p^2-m^2\right]}{2m^2
\left[\left(1+2a\right)p^2+2m^2\right]} \label{18}
\end{equation}
obey equations
\begin{equation}
\left(\pm i \check{p}+ m \right)\Pi_{\pm}=0 , \label{19}
\end{equation}
\begin{equation}
\Pi_{\pm}^2=\Pi_{\pm} ,~~~~\Pi_{+}\Pi_{-}=0 .\label{20}
\end{equation}
So, matrices $\Pi_{\pm}$ are projection matrices \cite{Fedorov}
and extract solutions of Eq. (12). For different mass states the
energy of particles, $p_0$ takes two values, Eq. (10). In
\cite{Kruglov2} (see also \cite{Kruglov}), we have obtained the
spin projection operators
\begin{equation}
P_{\pm 1/2}=\mp\frac{1}{2}\left(\sigma_p\pm\frac{1}{2}
\right)\left(\sigma_p^2-\frac{9}{4}\right) , \label{21}
\end{equation}
where the operator of the spin projections on the direction of the
momentum, $\textbf{p}$, is given by
\begin{equation}
\sigma_p=-\frac{i}{2|\textbf{p}|}\epsilon_{abc}\textbf{p}_a J_{bc}
. \label{22}
\end{equation}
We use the notation $|\textbf{p}| =\sqrt{p_1^2 +p_2^2+p_3^2}$, and
the generators of the Lorentz group in the 20-dimensional
reducible representation are \cite{Kruglov1}
\begin{equation}
J_{\mu \nu }=J_{\mu \nu }^{(1)}\otimes I_4+I_5 \otimes
J_{\mu\nu}^{(1/2)} ,
 \label{23}
\end{equation}
\begin{equation}
J_{\mu\nu}^{(1)}= \varepsilon^{\mu,\nu}-\varepsilon^{\nu,\mu} ,
 \label{24}
\end{equation}
\begin{equation}
J_{\mu\nu}^{(1/2)}= \frac{1}{4}\left( \gamma_\mu
\gamma_\nu-\gamma_\nu \gamma_\mu \right) .
 \label{25}
\end{equation}
The projection operator $P_{\pm 1/2}$ obeys the equation
\begin{equation}
\sigma_p P_{\pm 1/2}=\pm \frac{1}{2} P_{\pm 1/2} . \label{26}
\end{equation}
Eq. (26) guarantees that the operator $P_{\pm 1/2}$ extracts spin
projections $s=\pm 1/2$. The operators $\check{p}$ and $\sigma_p$
commute: $[ \check{p},\sigma_p]=0$ and, therefore, these operators
have the common eigenfunction in the momentum space.

With the aid of Eqs. (19), (26), we obtain the projection
operators for pure spin states in the form of matrix-dyads
(density matrices)
\begin{equation}
U_s(\pm p) \cdot \overline{U}_s(\pm p)=N P_{\pm 1/2}\Pi_\pm ,
\label{27}
\end{equation}
where $N$ is the normalization constant, and we imply the matrix
elements of matrix-dyads to be: $\left(U_s(\pm p) \cdot
\overline{U}_s(\pm p)\right)_{AB}=\left(U_s(\pm p)\right)_A
\left(\overline{U}_s(\pm p)\right)_B$. There is no summation over
spin indexes $s$ in Eq. (27).  The operator $U_s(\pm p)\cdot
\overline{U}_s(\pm p)$ extracts the solution of Eq. (12) as well
as the equation:
\begin{equation}
\sigma_p U_s(\pm p)=sU_s(\pm p) ,\label{28}
\end{equation}
where $s=\pm (1/2)$. The density matrices (27) are 20$\times$20
matrices describing the polarization of particles for pure spin
states.

For pure spin states we use the normalization:
\begin{equation}
\overline{U}_s(\pm p)U_r(\pm p)=\mp \delta_{sr}\frac{p^2}{m}
,~~~~\overline{U}_s( p)U_r(-p)=0 .\label{29}
\end{equation}
Eqs. (29) generalize the normalization condition (15) on the case
of states with definite spin projections. From Eq. (27), we obtain
the expression for matrix density summed over spin projections
\begin{equation}
\sum_s U_s(\pm p) \cdot \overline{U}_s(\pm p)=N_{\pm} \Pi_{\pm } .
\label{30}
\end{equation}
Here we took into consideration the relationship (see
Appendix)
\begin{equation}
 \check{p}\left( P_{1/2}+ P_{- 1/2}\right)=\check{p} .
\label{31}
\end{equation}
Taking the trace in both sides of Eq. (30), and using the equality
(see Appendix) $tr \Pi_{\pm }=2$, one finds
\begin{equation}
 \sum_s \overline{U}_s(\pm p)U_s(\pm p)=2N_{\pm} .
\label{32}
\end{equation}
If we compare Eq. (32) with the expression (29) summed over two
states $s=\pm 1/2$, we get
\begin{equation}
 N_{\pm}=\mp \frac{p^2}{m} ,
\label{33}
\end{equation}
for two values of $p^2$: $p^2=-m_1^2$ or $p^2=-m_2^2$.

\section{Quantization of Fermionic Fields with Two Mass States}

Following the general prescription \cite{Dirac}, we obtain from
Eq. (7) the momenta:
\begin{equation}
\pi (x)=\frac{\partial\mathcal{L}}{\partial(\partial_0\Psi
(x)}=\frac{i}{2}\overline{\Psi}(x)\alpha_4
 ,\label{34}
\end{equation}
\begin{equation}
\overline{\pi}(x)=\frac{\partial\mathcal{L}}{\partial(\partial_0
\overline{\Psi}(x))} =-\frac{i}{2}\alpha_4\Psi(x) . \label{35}
\end{equation}

We consider here the fields $\Psi(x)$, $\overline{\Psi} (x)$ as
independent ``coordinates". The Poisson bracket
$\left\{.,.\right\}^P$ of the ``coordinate" $\Psi (x)$ and the
momentum $\pi (x)$ is given by the standard equation
\begin{equation}
\left\{\Psi_M(\textbf{x},t),\pi_N (\textbf{y},t)\right\}^P=
\delta_{MN}\delta(\textbf{x}-\textbf{y}) ,
 \label{36}
\end{equation}
and indexes $M$, $N$ run 20 components. In the quantum field
theory, one should make the substitution for fermionic fields
\begin{equation}
\left\{\Psi_M(\textbf{x},t),\pi_N (\textbf{y},t)\right\}^P
\rightarrow  -i\left\{\Psi_M(\textbf{x},t),\pi_N
(\textbf{y},t)\right\} ,\label{37}
\end{equation}
where the $\left\{\Psi_M,\pi_N\right\}=\Psi_M\pi_N+\pi_N\Psi_M$ is
quantum anticommutator. Replacing Eqs. (34), (35) into Eqs. (36),
and taking into consideration Eq. (37), we arrive at
\begin{equation}
\left\{\Psi_M(\textbf{x},t),\left(\overline{\Psi}(\textbf{y},t)\alpha_4
\right)_N\right\}= 2\delta_{MN}\delta(\textbf{x}-\textbf{y})
.\label{38}
\end{equation}
From Eqs. (3), (9), (38), we obtain anticommutators of Dirac
spinors:
\begin{equation}
\left\{\psi_\mu(\textbf{x},t),\partial_0\overline{\psi}_\nu(\textbf{y},t)
\right\}=
-2i\frac{m}{a}\delta_{\mu\nu}\delta(\textbf{x}-\textbf{y})
,\label{39}
\end{equation}
\begin{equation}
\left\{\psi_\mu(\textbf{x},t),\psi^*_\nu(\textbf{y},t) \right\}=0
,\label{40}
\end{equation}
where $\partial_0=\partial/\partial t$, $\overline{\psi}(x)=\psi^+
(x)\gamma_4$. Commutation relations (39), (40) are different from
those of QED because the equations considered possess higher
derivatives, Eq. (1). In standard QED, anticommutator (40) does
not equal zero, and there is no the equation (39). The equation
like (39), including anticommutator of the field and its time
derivative, is typical for bosonic fields with the replacement of
anticommutators on commutators because of different statistics.
This is due to the fact that bosonic fields obey second order
equations as well as Eq. (1).

The density of the Hamiltonian (the energy density) may be found
from the equation
\begin{equation}
 {\cal H}=\pi (x)\partial_0\Psi (x)+\left(\partial_0\overline{\Psi}
 (x)\right)\overline{\pi} (x) -{\cal L}
.\label{41}
\end{equation}
Replacing Eqs. (34), (35) into Eq. (41), and taking into
consideration that ${\cal L}=0$ for fields satisfying Eq. (4), one
obtains
\begin{equation}
 {\cal H}=\frac{i}{2}\overline{\Psi}(x)\alpha_4\partial_0\Psi (x)
 -\frac{i}{2}\left(\partial_0\overline{\Psi}(x)\right)\alpha_4\Psi (x)
.\label{42}
\end{equation}
The density of the Hamiltonian (42) coincides with fourth
components, ${\cal H}=T_{44}$, of the the energy-momentum tensor
\cite{Kruglov}:
\[
T_{\mu\nu}=\frac{1}{2}\left(\partial_\nu \overline{\Psi}
(x)\right)\alpha_\mu \Psi (x)-\frac{1}{2} \overline{\Psi}
(x)\alpha_\mu \partial_\nu\Psi (x)
\]
\begin{equation}
=\frac{1}{2}\overline{\psi} (x)\gamma_\mu
\partial_\nu\psi (x)-\frac{1}{2}\left(\partial_\nu\overline{\psi} (x)\right)\gamma_\mu
\psi (x)+\frac{a}{2m}\left(\partial_\mu\overline{\psi}
(x)\right)\partial_\nu \psi (x) \label{43}
\end{equation}
\[
-\frac{a}{2m}\left(\partial_\mu\partial_\nu \overline{\psi}
(x)\right)\psi (x)+ \frac{a}{2m}\left(\partial_\nu \overline{\psi}
(x)\right)\partial_\mu\psi (x) -\frac{a}{2m}\overline{\psi}
(x)\partial_\nu
\partial_\mu\psi (x) .
\]

The electric current density  are given by \cite{Kruglov}
\[
j_\mu (x)=i\overline{\Psi }(x)\alpha_\mu \Psi(x)
\]
\vspace{-7mm}
\begin{equation} \label{44}
\end{equation}
\vspace{-7mm}
\[
= -i\overline{\psi }(x)\gamma_\mu \psi(x)+
\frac{ia}{m}\overline{\psi }(x)\partial_\mu \psi(x)
-\frac{ia}{m}\left(\partial_\mu\overline{\psi}(x)\right) \psi(x).
\]
This expression includes the usual Dirac current and Barut's
convective terms. The charge density follows from Eq. (44):
$j_0=-ij_4=\overline{\Psi }(x)\alpha_4 \Psi(x)$. Therefore, the
normalization condition (13) is the normalization on the charge.

In the second quantized theory the general solution of Eq. (4) may
be written as
\[
\Psi_\tau(x)=\sum_{s}\left[a_{\tau,s}\Psi^{(+)}_{\tau,s}(x) +
b^+_{\tau,s}\Psi^{(-)}_{\tau,s}(x)\right]
\]
\vspace{-7mm}
\begin{equation} \label{45}
\end{equation}
\vspace{-7mm}
\[
=\sum_{p,s}\frac{1}{\sqrt{2p_0 V}}\left[a_{\tau,p,s}U_{\tau,s}(
p)\exp( ipx)+ b^+_{\tau,p,s}U_{\tau,s}(- p)\exp(- ipx)\right] .
\]
Conjugated function $\overline{\Psi}(x)$ reads
\[
\overline{\Psi}_\tau(x)=\sum_{s}\left[a^+_{\tau,s}\overline{\Psi^{(+)}_{\tau,s}}(x)
+ b_{\tau,s}\overline{\Psi^{(-)}_{\tau,s}}(x)\right]
\]
\vspace{-7mm}
\begin{equation} \label{46}
\end{equation}
\vspace{-7mm}
\[
=\sum_{p,s}\frac{1}{\sqrt{2p_0
V}}\left[a^+_{\tau,p,s}\overline{U}_{\tau,s}( p)\exp(- ipx)+
b_{\tau,p,s}\overline{U}_{\tau,s}(- p)\exp( ipx)\right] ,
\]
where $a^+_{\tau,p,s}$, $a_{\tau,p,s}$ are the creation and
annihilation operators of particles, and $b^+_{\tau,p,s}$,
$b_{\tau,p,s}$ are the creation and annihilation operators of
antiparticles. The quantum number $\tau=1,2$ corresponds to two
mass states (10). Commutation relations are given by
\[
\{a_{\tau,p,s},a^+_{\tau',p',s'}\}=\delta_{ss'}\delta_{\tau\tau'}
\delta_{pp'} ,~~~\{a_{\tau,p,s},a_{\tau',p',s'}\}=0
,~~~\{a^+_{\tau,p,s},a^+_{\tau',p',s'}\}=0 ,
\]
\begin{equation}
\{b_{\tau,p,s},b^+_{\tau',p',s'}\}=\delta_{ss'}\delta_{\tau\tau'}
\delta_{pp'} ,~~~\{b_{\tau,p,s},b_{\tau',p',s'}\}=0
,~~~\{b^+_{\tau,p,s},b^+_{\tau',p',s'}\}=0 , \label{47}
\end{equation}
\[
\{a_{\tau,p,s},b_{\tau',p',s'}\}=\{a_{\tau,p,s},b^+_{\tau',p',s'}\}
=\{a^+_{\tau,p,s},b_{\tau',p',s'}\}=\{a^+_{\tau,p,s},b^+_{\tau',p',s'}\}=0
.
\]
With the help of Eqs. (45)-(47), and normalization condition (13),
one obtains from Eq. (42) the energy of particles-antiparticles
fields
\begin{equation}
H=\int {\cal H}d^3 x=\sum_{\tau,p,s}p_0\left(a^+_{\tau,p,s}
a_{\tau,p,s}-b_{\tau,p,s} b^+_{\tau,p,s}\right) . \label{48}
\end{equation}
Like QED, we find from Eqs. (45)-(47) commutation relations for
fields $\Psi_\tau(x)$, $\overline{\Psi}_\tau(x)$:
\begin{equation}
\{\Psi_{\tau M(x)},\Psi_{\tau N}(x')\}=\{\overline{\Psi}_{\tau
M}(x), \overline{\Psi}_{\tau N}(x')\ =0, \label{49}
\end{equation}
\begin{equation}
\{\Psi_{\tau M}(x),\overline{\Psi}_{\tau N}(x')\}=K_{\tau
MN}(x,x'), \label{50}
\end{equation}
\[
K_{\tau MN}(x,x')=K^+_{\tau MN}(x,x')+K^-_{\tau MN}(x,x') ,
\]
\begin{equation}
K^+_{\tau MN}(x,x')=\sum_{s}\left(\Psi^{(+)}_{\tau,s}(x)\right)_M
\left(\overline{\Psi^{(+)}_{\tau,s}}(x')\right)_N  ,\label{51}
\end{equation}
\[
K^-_{\tau MN}(x,x')=\sum_{s}\left(\Psi^{(-)}_{\tau,s}(x)\right)_M
\left(\overline{\Psi^{(-)}_{\tau,s}}(x')\right)_N .
\]
From Eqs. (45), (46), one finds
\[
K^+_{\tau MN}(x,x')=\sum_{p,s}\frac{1}{2p_0 V}\left(U_{\tau,s}(
p)\right)_M\left(\overline{U}_{\tau,s}(p)\right)_N\exp [ip(x-x')]
,
\]
\vspace{-7mm}
\begin{equation} \label{52}
\end{equation}
\vspace{-7mm}
\[
K^-_{\tau MN}(x,x')=\sum_{p,s}\frac{1}{2p_0 V}\left(U_{\tau,s}(
-p)\right)_M\left(\overline{U}_{\tau,s}(-p)\right)_N\exp
[-ip(x-x')] .
\]
Taking into account Eqs. (30), (33), and the relation
$p^2=-m_\tau^2$, we obtain from Eqs. (52) functions as follows:
\[
K^+_{\tau MN}(x)=\sum_{p}\left( \frac{i\check{p}\left(m-
i\check{p}\right)
\left[\check{p}^2+\left(1+2a\right)m_\tau^2-m^2\right]m_\tau^2}{4p_0Vm^3
\left[2m^2-\left(1+2a\right)m_\tau^2\right]}\right)_{MN}\exp (ipx)
\]
\begin{equation}
=\left( \frac{\alpha_\mu \partial_\mu\left(m-\alpha_\nu
\partial_\nu\right)
\left[\left(1+2a\right)m_\tau^2-m^2-(\alpha_\mu
\partial_\mu)^2\right]m_\tau^2}{2m^3
\left[2m^2-\left(1+2a\right)m_\tau^2\right]}\right)_{MN}
\label{53}
\end{equation}
\[
\times\sum_{p}\frac{1}{2p_0V}\exp(ipx) ,
\]
\[
K^-_{\tau MN}(x)=\sum_{p}\left( \frac{i\check{p}\left(m+
i\check{p}\right)
\left[\check{p}^2+\left(1+2a\right)m_\tau^2-m^2\right]m_\tau^2}{4p_0Vm^3
\left[2m^2-\left(1+2a\right)m_\tau^2\right]}\right)_{MN}\exp
(-ipx)
\]
\begin{equation}
=-\left( \frac{\alpha_\mu \partial_\mu\left(m-\alpha_\nu
\partial_\nu\right)
\left[\left(1+2a\right)m_\tau^2-m^2-(\alpha_\mu
\partial_\mu)^2\right]m_\tau^2}{2m^3
\left[2m^2-\left(1+2a\right)m_\tau^2\right]}\right)_{MN}
\label{54}
\end{equation}
\[
\times\sum_{p}\frac{1}{2p_0V}\exp (-ipx) .
\]
Using the singular functions \cite{Ahieser}
\[
\Delta_+(x)=\sum_{p}\frac{1}{2p_0V}\exp
(ipx),~~~~\Delta_-(x)=\sum_{p}\frac{1}{2p_0V}\exp (-ipx),
\]
\[
\Delta_0 (x)=i\left(\Delta_+(x)-\Delta_-(x)\right),
\]
we obtain from Eqs. (51), (53), (54)
\[
K_{\tau MN}(x)=-i\left( \frac{\alpha_\mu
\partial_\mu\left(m-\alpha_\nu
\partial_\nu\right)
\left[\left(1+2a\right)m_\tau^2-m^2-(\alpha_\mu
\partial_\mu)^2\right]m_\tau^2}{2m^3
\left[2m^2-\left(1+2a\right)m_\tau^2\right]}\right)_{MN}
\]
\vspace{-6mm}
\begin{equation} \label{55}
\end{equation}
\vspace{-6mm}
\[
\times\Delta_0 (x) .
\]
There is no summation in index $\tau$ in Eqs. (49)-(55). As in
QED, due to the properties of the function $\Delta_0 (x)$,
anticommutator $\{\Psi_M (x),\overline{\Psi}_N(x')\}$ equals zero
if the points $x$ and $x'$ are separated by the space-like
interval $(x-x')>0$. For equal times, $t=t'$, one has $\{\Psi_M
(\textbf{x},0),\overline{\Psi}_N(\textbf{x}')\}=K_{\tau
MN}(\textbf{x}-\textbf{x}',0)$, where the function $K_{\tau
MN}(\textbf{x}-\textbf{x}',0)$ may be obtained from Eq. (55) with
the help of equalities
\begin{equation}
\partial_0^{2n}\Delta_0 (x)|_{t=0}=0 ,~~~~\partial_m^{n}\Delta_0
(x)|_{t=0}=0 ,~~~~\partial_0 \Delta_0 (x)|_{t=0}= \delta
(\textbf{x}) , \label{56}
\end{equation}
where $n=1,2,3,...$. It is easy to verify, using Eq. (17), that
the equations
\begin{equation}
\left(\alpha_\mu\partial_\mu +m\right)K^-_{\tau}(x)=0 ,
~~~~\left(\alpha_\mu\partial_\mu +m\right)K^+_{\tau}(x)=0
 \label{57}
\end{equation}
are valid. The chronological pairing of the operators in our
formalism are given by \cite{Ahieser}
\[
\Psi^a_{\tau M}(x)\overline{\Psi}^a_{\tau N}(y)=K^c_{\tau MN}(x-y)
\]
\vspace{-6mm}
\begin{equation} \label{58}
\end{equation}
\vspace{-6mm}
\[
=\theta\left(x_0 -y_0\right)K^+_{\tau MN}(x-y)-\theta\left(y_0
-x_0\right)K^-_{\tau MN}(x-y) ,
\]
where $\theta(x)$ is the well known theta-function. With the aid
of Eqs. (53), (54), one finds
\[
\Psi^a_{\tau M}(x)\overline{\Psi}^a_{\tau N}(y)
\]
\vspace{-6mm}
\begin{equation} \label{59}
\end{equation}
\vspace{-6mm}
\[
=\left( \frac{\alpha_\mu\partial_\mu
\left(m-\alpha_\nu\partial_\nu\right)
\left[\left(1+2a\right)m_\tau^2-m^2-(\alpha_\mu
\partial_\mu)^2\right]m_\tau^2}{2m^3
\left[2m^2-\left(1+2a\right)m_\tau^2\right]}\right)_{MN}\Delta_c
(x-y) ,
\]
and the function $\Delta_c (x-y)$ is given by
\begin{equation}
\Delta_c (x-y)=\theta\left(x_0
-y_0\right)\Delta_+(x-y)+\theta\left(y_0 -x_0\right)\Delta_-(x-y)
. \label{60}
\end{equation}

\section{Conclusion}

In the formalism of the first order, we have obtained the
projection operators extracting states with definite energy and
spin projections of the generalized Dirac equation, describing
particles with spin 1/2 and two mass states. The density matrix
was found for pure spin states. The canonical quantization was
performed and anticommutators of 20-component fields were obtained
in this formalism. So, FOGDE is convenient for a consideration of
the conserving currents as well as for quantization of fields. The
density matrix and the chronological pairing of the operators
found allow us to calculate different quantum possesses in the
formalism of the first order.

\section{Appendix}

For convenience, we write down some matrices entering the matrix
density and singular functions. From Eq. (5), one finds
\begin{equation}
\check{p}\equiv p_\nu\alpha _\nu
=I^{(0)}\otimes\hat{p}+I^{(1)}\otimes I_4
 ,\label{61}
\end{equation}
where $\hat{p}\equiv p_\nu\gamma_\nu$, and
\begin{equation}
I^{(0)}\equiv \varepsilon^{0,0} ,~~~~I^{(1)}\equiv
p_\nu\left(\varepsilon ^{\nu,0 }+ a\varepsilon ^{0,\nu }\right)
,~~~~I^{(2)}\equiv p_\mu p_\nu \varepsilon ^{\mu,\nu } .
\label{62}
\end{equation}
With the aid of Eq. (6), we obtain matrices as follows:
\begin{equation}
\check{p}^2 =\left(1+a\right)p^2I^{(0)}\otimes
I_4+I^{(1)}\otimes\hat{p}+aI^{(2)}\otimes I_4 ,\label{63}
\end{equation}
\begin{equation}
\check{p}^3
=\left(1+2a\right)p^2I^{(0)}\otimes\hat{p}+\left(1+a\right)p^2I^{(1)}\otimes
I_4+aI^{(2)}\otimes \hat{p} ,\label{64}
\end{equation}
\begin{equation}
\check{p}^4 =\left(1+3a+a^2\right)p^4I^{(0)}\otimes
I_4+\left(1+2a\right)p^2I^{(1)}\otimes\hat{p}+a\left(1+a\right)p^2I^{(2)}\otimes
I_4 ,\label{65}
\end{equation}
\[
\check{p}^5
=\left(1+4a+3a^2\right)p^4I^{(0)}\otimes\hat{p}+\left(1+3a+a^2\right)p^4I^{(1)}\otimes
I_4
\]
\vspace{-7mm}
\begin{equation} \label{66}
\end{equation}
\vspace{-7mm}
\[
 +a\left(1+2a\right)p^2I^{(2)}\otimes \hat{p} ,
\]
\[
\check{p}^6 =\left(1+5a+6a^2+a^3\right)p^6I^{(0)}\otimes
I_4+\left(1+4a+3a^2\right) p^4I^{(1)}\otimes\hat{p}
\]
\vspace{-7mm}
\begin{equation} \label{67}
\end{equation}
\vspace{-7mm}
\[
+a\left(1+3a+a^2\right)p^4I^{(2)}\otimes I_4 .
\]
The matrices (62) obey the equations:
\[
I^{(0)2}=I^{(0)} ,~~~~I^{(0)}I^{(1)}+I^{(1)}I^{(0)}=I^{(1)},~~~~
I^{(2)2}=p^2I^{(2)} ,
\]
\vspace{-6mm}
\begin{equation} \label{68}
\end{equation}
\vspace{-6mm}
\[
I^{(0)}I^{(2)}=I^{(2)}I^{(0)}=0
,~~~~I^{(2)}I^{(1)}+I^{(1)}I^{(2)}=p^2I^{(1)}.
\]

\end{document}